The detection of "hot regions" in the geography of science –

A visualization approach by using density maps

Lutz Bornmann$, Ludo Waltman§

$ Max Planck Society, Hofgartenstr. 8, 80539 Munich, Germany

§ Centre for Science and Technology Studies, Leiden University, Leiden, The Netherlands

Corresponding author:

Lutz Bornmann, bornmann@gv.mpg.de

**Abstract**

Spatial scientometrics has attracted a lot of attention in the very recent past. The visualization methods (density maps) presented in this paper allow for an analysis revealing regions of excellence around the world using computer programs that are freely available. Based on Scopus and Web of Science data, field-specific and field-overlapping scientific excellence can be identified in broader regions (worldwide or for a specific continent) where high quality papers (highly cited papers or papers published in *Nature* or *Science*) were published. We used a geographic information system to produce our density maps. We also briefly discuss the use of Google Earth.

# 1 Introduction

Spatial scientometrics has attracted a lot of attention in the very recent past. In *Nature News* van Noorden (2010) discusses urban regions producing the best research and whether their success can be replicated. In a literature overview, Frenken et al. (2009) identified some descriptive studies which investigated differences among regions worldwide, in terms of their publication output and citations. For example, Matthiessen, Schwarz, and Find (2002) analyzed the strength, interrelations, and nodality of global research centers. The study was based on publication and citation data for the period from 1997 to 1999. The publications were written by authors located in the 40 largest "greater" urban regions of the world. Three types of regions were identified on the basis of levels of output and citation impact. According to this categorization, group 1 combines very high impact (measured as average citation rate of their publications) with very high total volume. This applied to New York, Boston and the San Francisco Bay Area. Group 2, characterized by high impact and medium level of total number of citations, comprises San Diego, Seattle, St Louis, Baltimore, Mannheim-Heidelberg, Cambridge, Basel-Mulhouse-Freiburg and Geneva-Lausanne. Group 3, with moderate impact and high volume, contains seven centres – London, Los Angeles, Philadelphia, Paris, Amsterdam-The Hague-Rotterdam-Utrecht, the Tokyo Bay Area and Osaka-Kobe-Kyoto (see here also Matthiessen, Schwarz, & Find, 2010).

In spatial scientometrics, the visualization of geographical data is receiving more and more attention. Boyack, Klavans, Paley, and Börner (2007) presented an interactive visualization in which a traditional science map and a geographical map of science were linked to each other. Leydesdorff and Persson (2010) explored the use of tools such as Google Maps and Google Earth for producing geographical maps of science. Bornmann, Leydesdorff, Walch-Solimena, and Ettl (in preparation) presented new methods to map centers of excellence around the world using freely available computer programs. By colorizing cities in dependency on the numbers of excellent papers, their maps provide visualizations where cities with a high (or low) output of excellent papers can be found. To demonstrate their approach, Bornmann, Leydesdorff, Walch-Solimena, and Ettl (in preparation) showed maps for excellent papers in neuroscience, physics & astronomy and social sciences as well as for *Nature* and *Science* articles. The present study is intended to follow their approach in general, but to change the focus from mapping of single data points (i.e., cities) to a more "sliding" visualization of regions. In our opinion this visualization, which is based on density maps, can serve as a useful complement to the maps proposed by Bornmann, Leydesdorff, Walch-Solimena, and Ettl (in preparation). The advantages of the new visualization are that (1) the few broader regions characterized by an excellent paper output can be identified (worldwide or for a specific continent) and (2) raw address data can be used, without using urban groupings (as Matthiessen, et al., 2010, did) to show results at the urban level. Because of the second advantage, no (arbitrary) aggregation of addresses is necessary.

In the following (1) we explain how the data sets that are needed can be produced and how the density maps can be calculated. (2) We present some examples of density maps and (3) discuss finally the advantages and disadvantages of this type of visualization. We have put the most important material to produce the maps online at www.ludowaltman.nl/density_map/. On this web page we have also made available a document that gives detailed instructions on how to produce the density maps.

## 2    Methods

*The data sets that have been used*

Our mapping approach is based on the visualization of scientific excellence. In the present study we follow the classification of Thomson Reuters (Philadelphia, PA, USA) and the National Science Board (2010) and focus on the top 1% of papers published in 2007 worldwide with a fixed citation window of three years (from 2007 up to the date of research, at the end of 2010 and in March 2011, respectively). Scopus is used as data base for this study since it allows one to easily select all papers published in a broader field (e.g., physics & astronomy).

In the following, the procedure to gather the excellent papers (more precisely: the cities of the authors having published the top 1% most highly cited papers) in a certain field is described in short. A detailed description of how to produce the data sets for our mapping approach can be found in Bornmann, Leydesdorff, Walch-Solimena, and Ettl (in preparation) and on the Internet: http://www.leydesdorff.net/maps/ (Leydesdorff & Persson, 2010) and http://www.leydesdorff.net/mapping_excellence/. Bornmann, Leydesdorff, Walch-Solimena, and Ettl (in preparation) give also detailed instructions on how to check the data quality. For example, geocoding does not always work perfectly and sometimes yields incorrect geographical coordinates (see here Waltman, Tijssen, & van Eck, 2011).

The procedure starts with a search for all papers published within one field (e.g., physics & astronomy). By sorting the search results by citation counts in decreasing order, the 1% of papers at the top of the Scopus list can be marked. The selected documents are exported and processed by four different programs (e.g., scopcity.exe). The programs extract the city names given by the co-authors of a paper into a text file. If there is more than one co-author of a paper with an identical address, this leads to a single address (or a single city occurrence). If the scientists are affiliated with different departments within the same institution, this leads to two addresses or two city occurrences, respectively. The city names in the text file are copied-and-pasted into the geocoding service at http://www.gpsvisualizer.com/geocoder/. This service provides for each city name the corresponding geographical coordinates (i.e., latitude and longitude). The output of the geocoding service is processed by a program for the last time. The final output is a text file that can be used as the coordinates file to produce the density maps.

We note that our processing of the address data involves some choices that may be subject to criticism. For instance, one may argue that affiliations with different departments within the same institution should yield only one city occurrence. Also, one may argue that publications with multiple addresses should be counted fractionally. A publication with addresses in three different cities would then count as one-third of a publication for each of the three cities. In such a fractional counting approach, each publication has a total weight of one.

*The calculation of the density maps*

Visualizations similar to those proposed in this study are also being used in other scientific fields. Shi (2010) provides references for ecology, criminology, public health, and epidemiology. Thus the contribution of this study is not to propose an entirely new type of visualization but rather to introduce an existing type of visualization into the field of spatial scientometrics. Our proposed visualization is inspired by the density maps offered by the VOSviewer software for bibliometric mapping (van Eck & Waltman, 2010). Examples of the use of these density maps can be found in van Eck and Waltman (2007) and Waaijer, van Bochove, and van Eck (2011). The main difference with this earlier work is that in the present study we work with geographical maps rather than with maps of some abstract space of, for instance, documents or keywords.

We use the following methodology to make our density maps. Suppose we want to make a density map of an area with southern boundary $\varphi_S$, northern boundary $\varphi_N$, western boundary $\lambda_W$, and eastern boundary $\lambda_E$, where $\varphi_S$ and $\varphi_N$ are latitudes and $\lambda_W$ and $\lambda_E$ are longitudes. We first construct a raster (or grid) that exactly covers the area of interest. To do so we need to specify the number of rows $n_R$ and the number of columns $n_C$ of the raster. The larger the number of rows and columns, the more accurate the density map will be. The raster has $n_R \times n_C$ cells. The center of the cell in the $i$th row and the $j$th column of the raster has latitude $\varphi(i)$ and longitude $\lambda(j)$, which are given by

$$\varphi(i) = \varphi_S + (i - 0.5)\frac{\varphi_N - \varphi_S}{n_R} \quad \text{and} \quad \lambda(j) = \lambda_W + (j - 0.5)\frac{\lambda_E - \lambda_W}{n_C}.$$

Suppose we have a large number of publications for which we know the geographical coordinates of the authors. Suppose there are *m* different pairs of coordinates. The *k*th pair of coordinates is given by a latitude $\varphi_k$ and a longitude $\lambda_k$. Each pair of coordinates has a weight $w_k$ indicating the number of publications associated with that pair of coordinates. Based on these data, we calculate for each cell in our raster the density of publications. The way in which this is done is similar to the statistical technique of kernel density estimation (e.g., Scott, 1992). For the cell in the *i*th row and the *j*th column of our raster, we calculate the density of publications as

$$D(i,j) = \frac{1}{h^2} \sum_{k=1}^{m} w_k K\left(\frac{d(\varphi_k, \lambda_k, \varphi(i), \lambda(j))}{h}\right),$$

where $d(\varphi_k, \lambda_k, \varphi(i), \lambda(j))$ denotes the great-circle distance from $\varphi_k$ and $\lambda_k$ to $\varphi(i)$ and $\lambda(j)$, *h* denotes the kernel width parameter, and *K* denotes the kernel function. The kernel width parameter *h* determines the smoothness of the density map. Choosing an appropriate value for this parameter is crucial for a satisfactory map. Some trial and error is usually needed to find a good value. The kernel function *K* that we use is given by

$$K(u) = 0.5 \exp(-u).$$

Our kernel function is somewhat unusual. Kernel density estimation typically uses a Gaussian kernel function or a kernel function with a shape similar to a Gaussian function (e.g., Scott, 1992). The density maps produced by the VOSviewer software for bibliometric mapping are also based on a Gaussian kernel function (van Eck & Waltman, 2010). However, in the case of geographical density maps based on address data from publications, some preliminary tests that we performed revealed that a Gaussian kernel function does not give completely satisfactory results. This seems to be caused by the highly irregular nature of the address data, in particular the characteristic that a relatively small set of addresses covers a large majority of the publications. We found that the above kernel function, which unlike a Gaussian kernel function is completely convex, yields more satisfactory results.

After calculating the density of publications for each cell in our raster, we determine the color of each cell. In this paper, we use a color scheme consisting of the colors white, green, yellow, and red. White indicates the lowest publication density: no publications at all. Yellow means twice as many publications as green, and red means five (or more) times as many publications as green. Thus red indicates the highest publication density. The color coding is done in such a way that colors in one map can be compared with colors in another map.

As a final step, we use a geographic information system (GIS) to display the density map. The GIS shows the raster of colors that we have constructed. On top of the raster, it shows the borders of countries and the locations of publications. The GIS that we use in this paper is Quantum GIS, an open source GIS available at www.qgis.org.

## 3    Results

To demonstrate the method to create density maps we produced two field-specific and two field-overlapping maps focusing on the European region. For the given results, a raster with 500 rows and 1000 columns and a kernel width parameter of 100 km were used. Figure 1 shows the location of authors in Europe having published highly cited papers in biochemistry, genetics, & molecular biology. The map is based on the top 1% of articles published in 2007 in a journal of the Scopus journal set biochemistry, genetics, & molecular biology. With the search string "subjarea(BIOC) and pubyear is 2007 and doctype(ar)" in the advanced search field of Scopus all papers with the document type "article" were retrieved which were published in 2007 within this Scopus journal set. On March 30, 2011 this search resulted in 171,663 papers. The search was restricted to articles (as document types) since (1) the method proposed here is intended to identify excellence at the research front and (2) different document types have different expected citation rates, possibly resulting in non-comparable data sets. By sorting the search results by citation counts in decreasing order (citation window: from 2007 to the date of search), the 1% of papers at the top of the Scopus list could be marked. On the date of search, 1,740 papers with at least 78 citations each (gathered between 2007 and the date of search) were marked as the list of the top 1% papers in biochemistry, genetics, & molecular biology.

On the map (Figure 1) the individual locations of the authors having published the top 1% papers (the small circles on the map) and the geographical regions with the different density colors (red, yellow, green, and white) are visible. For our mapping approach, we decided to also draw densities on sea and ocean areas, as the maps look better when densities are also allowed to cover these areas (although there can not be any author of an excellent paper). If we focus on geographical regions with a higher concentration of excellent papers (yellow colored regions), these regions are: (1) around Paris, (2) some western areas of Germany, as well as (3) broad areas in Belgium and the Netherlands. The highest concentration of excellent papers can be found in the region London – Cambridge – Oxford. These results are more or less in accordance with those published so far (see the Introduction section).

Figure 2 shows the corresponding map for physics & astronomy. All in all 146,081 articles were published in 2007 in this journal set worldwide; the top 1% are those 1,501 papers which received at least 44 citations each between 2007 and the date of research (March 29, 2011). Here again, we see a high density of excellent papers in the region London – Cambridge – Oxford, around Paris, and in Belgium and the Netherlands. Comparing the physics & astronomy and the biochemistry, genetics, & molecular biology maps (colors in one map can directly be compared with colors in another map, see above), it can be observed that Germany plays a more prominent role in the former than in the latter: Broad regions in the western and southern part show a relatively high concentration of excellent papers in physics & astronomy with yellow coloring.

As a final step in the analyses for this study we leave the focus of the density mapping on specific fields and present the results of a field-overlapping analysis of excellent papers. For this, we downloaded from Web of Science in two separate steps the bibliographic data of all articles published in 2007 and 2009 in the high-impact journals *Nature* and *Science*. This analysis is inspired by maps of Luis Bettencourt and Jasleen Kaur (Indiana University in Bloomington) published on www.nature.com/news/specials/cities/best-cities.html. The authors analyzed city addresses appearing in *Science*, *Nature*, and *Proceedings of the National Academy of Sciences* in 1989, 1999, and 2009.

We downloaded the bibliographic data of all 1,604 *Nature* and *Science* articles published in 2007 (date of research: October 27, 2010) and of all 1,643 *Nature* and *Science* articles published in 2009 (date of research: November 23, 2010). The maps based on Web of Science data are presented in Figure 3 (year: 2007) and Figure 4 (year: 2009), respectively. In agreement to the field-specific maps presented above, the highest densities are visible around London – Cambridge – Oxford. When comparing the two maps (for 2007 and 2009), it can be concluded that the overall contribution of the UK to these two journals (relative to the total publication output of the journals) has decreased between 2007 and 2009.

We managed to get our density map approach working in Google Earth as well (Figure 5). The procedure is similar to what we have described above for traditional GIS software, only the technical details are somewhat different. The discrete nature of the density raster is much less visible than in traditional GIS software, especially when one zooms in. When one zooms in, more and more locations of authors having published excellent papers gradually become visible. On www.ludowaltman.nl/density_map/, we offer the possibility to download four KMZ files, one for each of our data sets (biochemistry, genetics, & molecular biology, physics & astronomy, and two data sets with *Nature* and *Science* papers). These KMZ files can be opened in Google Earth.

# 4 Discussion

The methods presented in this paper allow for an analysis revealing regions of excellence around the world using computer programs that are freely available. Based on Scopus data, field-specific excellence can be identified in broader regions where highly cited papers were published. Thus, the methods as described above are especially oriented on broader excellence regions and do not allow for the identification of research institutions or single cities on the map where the authors of excellent papers are located. The identification of broader regions receives its importance against the backdrop of the so called reverse N-effect: "More competitors (here: prolific scientists) working within the same region produce better results (here: a higher output of excellent papers). Due to a systematic spatial bias in the interaction prolific researchers thus favor other prolific researchers for cooperation in research located in physical closeness" (Bornmann, et al., in preparation). Although our results have shown that the overall contribution of the "hot region" in southern UK to *Nature* and *Science* has decreased over two years, it is expected in general that a high density of excellent researchers generates a fruitful environment to produce future excellent research (see here also Bornmann, de Moya-Anegón, & Leydesdorff, 2010).

Despite the advantages of our mapping approach to visualize excellence in science, we recognize the limitations of using bibliometric data (see here Bornmann, et al., in preparation). (1) It is not guaranteed that the addresses listed on a publication reflect the locations where the reported research was conducted. (2) Each literature data base covers only a part of all published contributions worldwide. (3) There are different publication and citation behaviors in different scientific fields which make the direct comparison of maps problematic. Compared to natural sciences, social sciences' and humanities' research results, for instance, are more frequently published in the scientists' native languages and as books (Iribarren-Maestro, Lascurain-Sánchez, & Sanz-Casado, 2009). (5) No standard technique exists for the subject classification of articles (see here Bornmann & Daniel, 2008; Bornmann, Mutz, Neuhaus, & Daniel, 2008).

Besides these general problems connected to bibliometric data there are some problems inherent to the specific approach proposed here:

1) A significant amount of manual parameter tuning is necessary in order to obtain insightful density maps.

2) The maps are not useful for small data sets. There should be at least a few hundred addresses available.

3) The correct interpretation of a density map may not always be obvious to everyone. Some people may be confused because areas without a university (and even sea and ocean areas) may still have a green, yellow, or red color (due to the interpolation that takes place in the calculation of the density).

## Acknowledgement

We would like to thank Loet Leydesdorff, Amsterdam School of Communications Research, for his help with the geocoding of the address data.

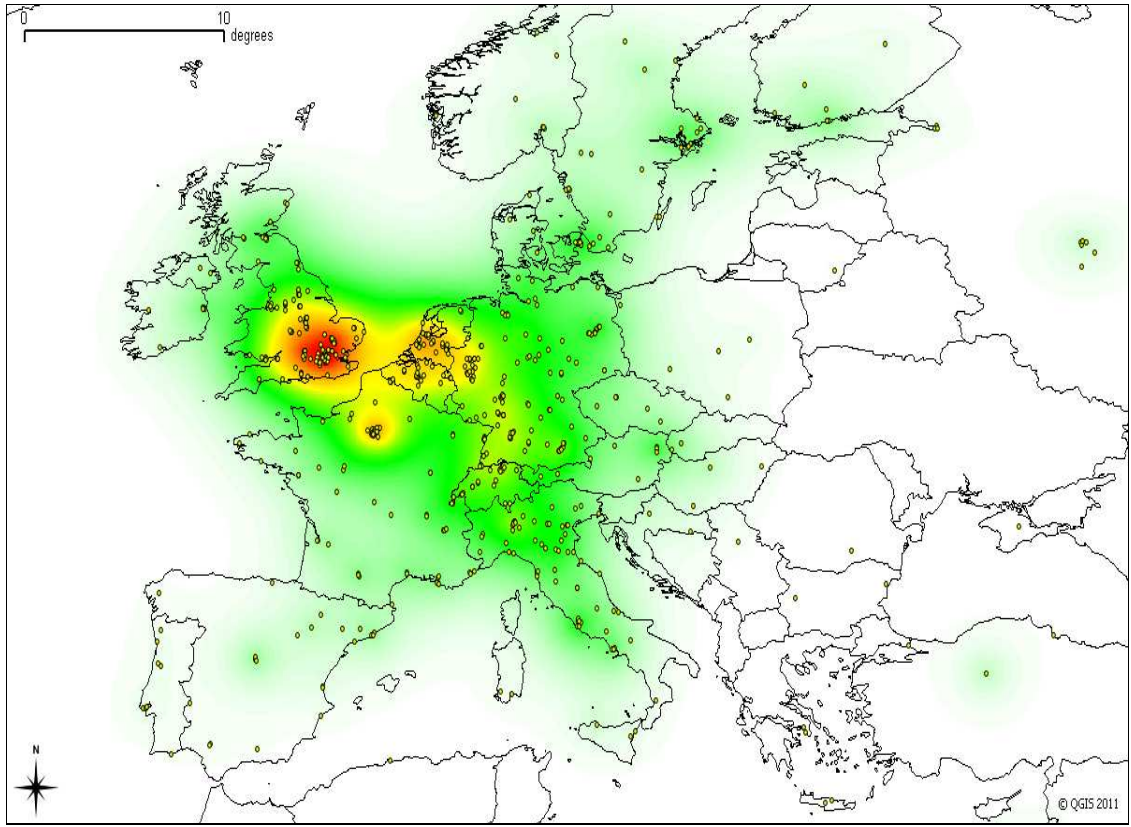

Figure 1. Density map of authors in Europe having published highly cited biochemistry, genetics & molecular biology papers in 2007 (searched in Scopus). This figure appears in color on the Web (PDF and HTML of this paper), but is not reproduced in color in the printed version.

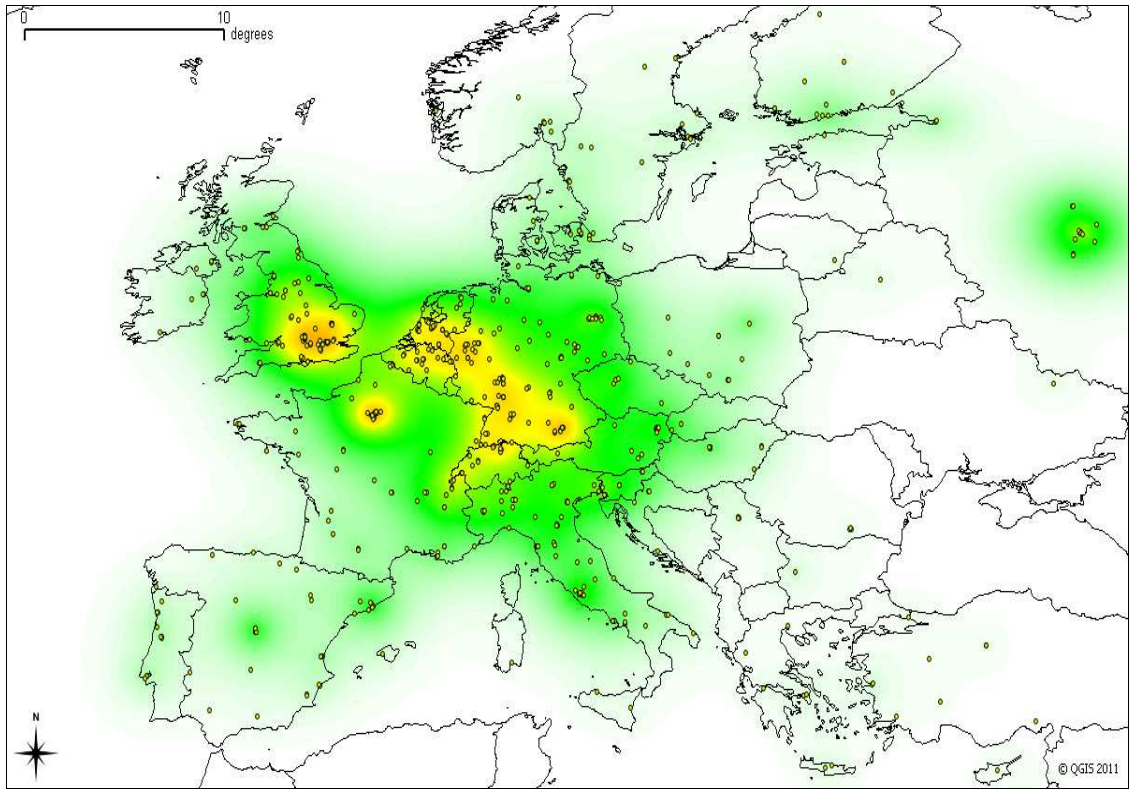

Figure 2. Density map of authors in Europe having published highly cited physics & astronomy papers in 2007 (searched in Scopus). This figure appears in color on the Web (PDF and HTML of this paper), but is not reproduced in color in the printed version.

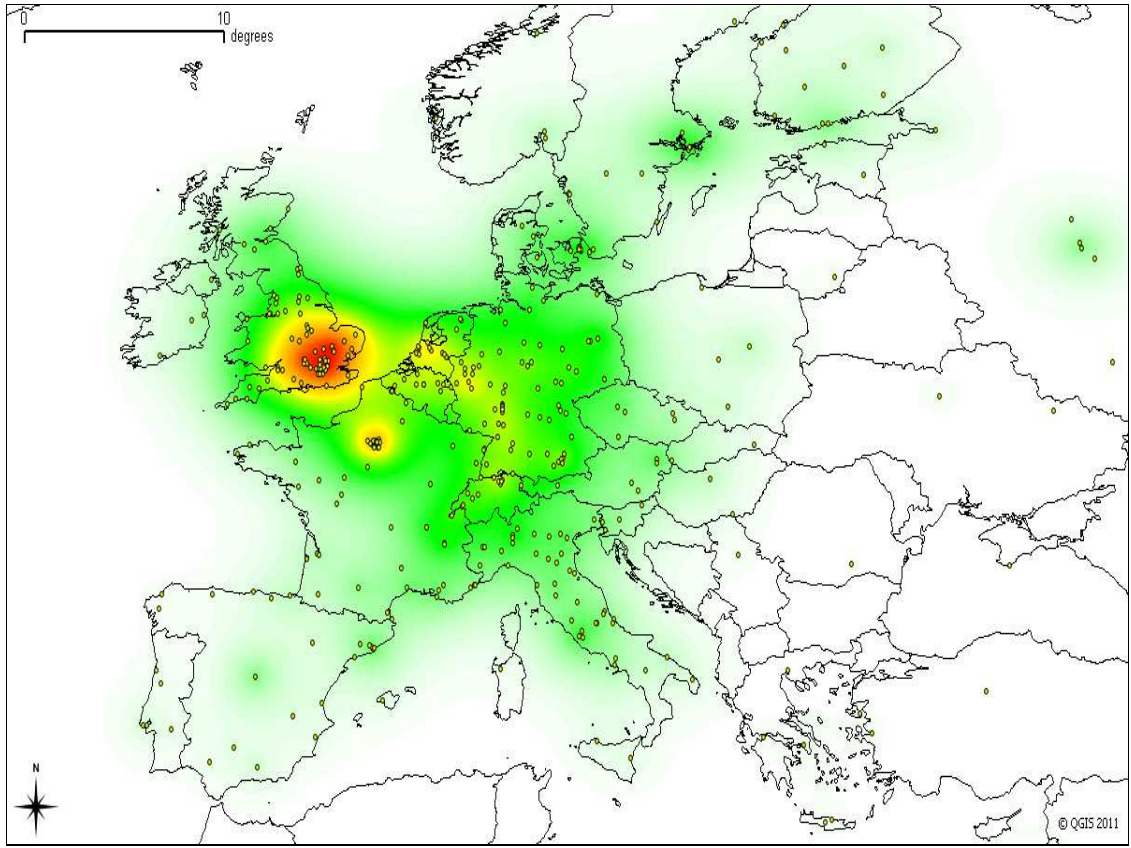

Figure 3. Density map of authors in Europe having published *Nature* or *Science* articles in 2007 (searched in Web of Science). This figure appears in color on the Web (PDF and HTML of this paper), but is not reproduced in color in the printed version.

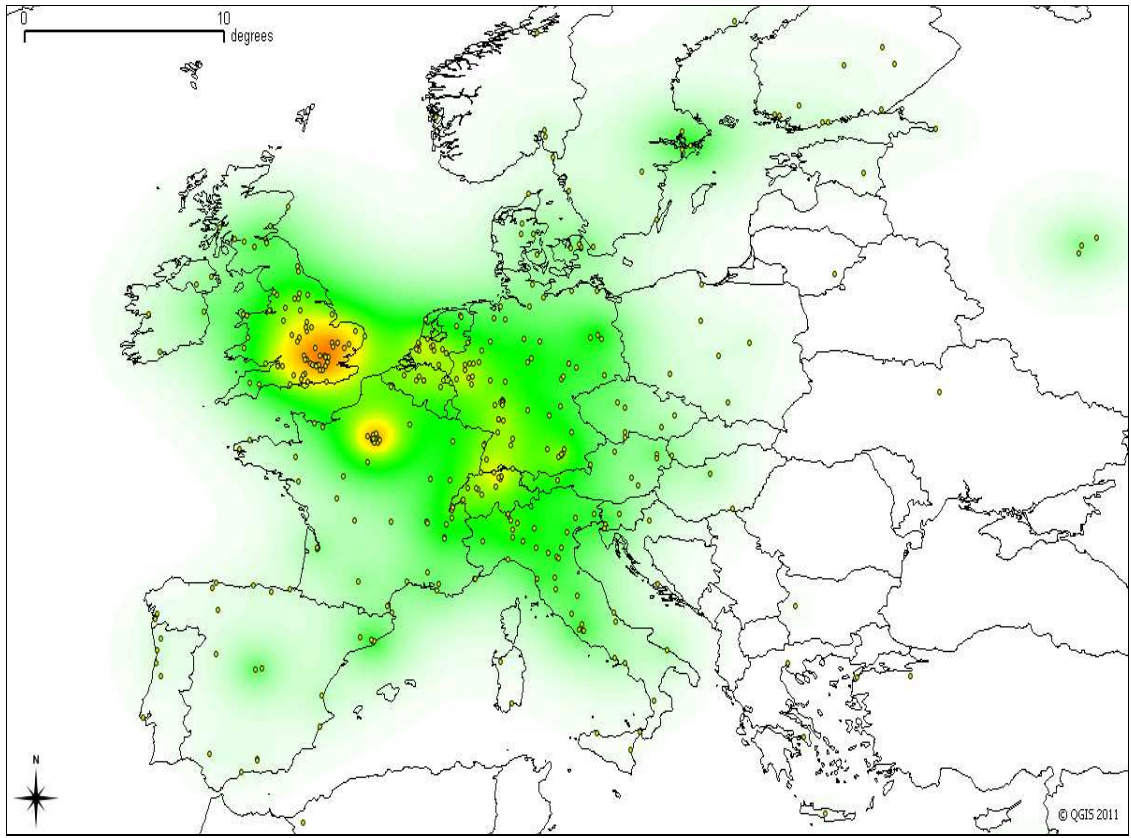

Figure 4. Density map of authors in Europe having published *Nature* or *Science* articles in 2009 (searched in Web of Science). This figure appears in color on the Web (PDF and HTML of this paper), but is not reproduced in color in the printed version.

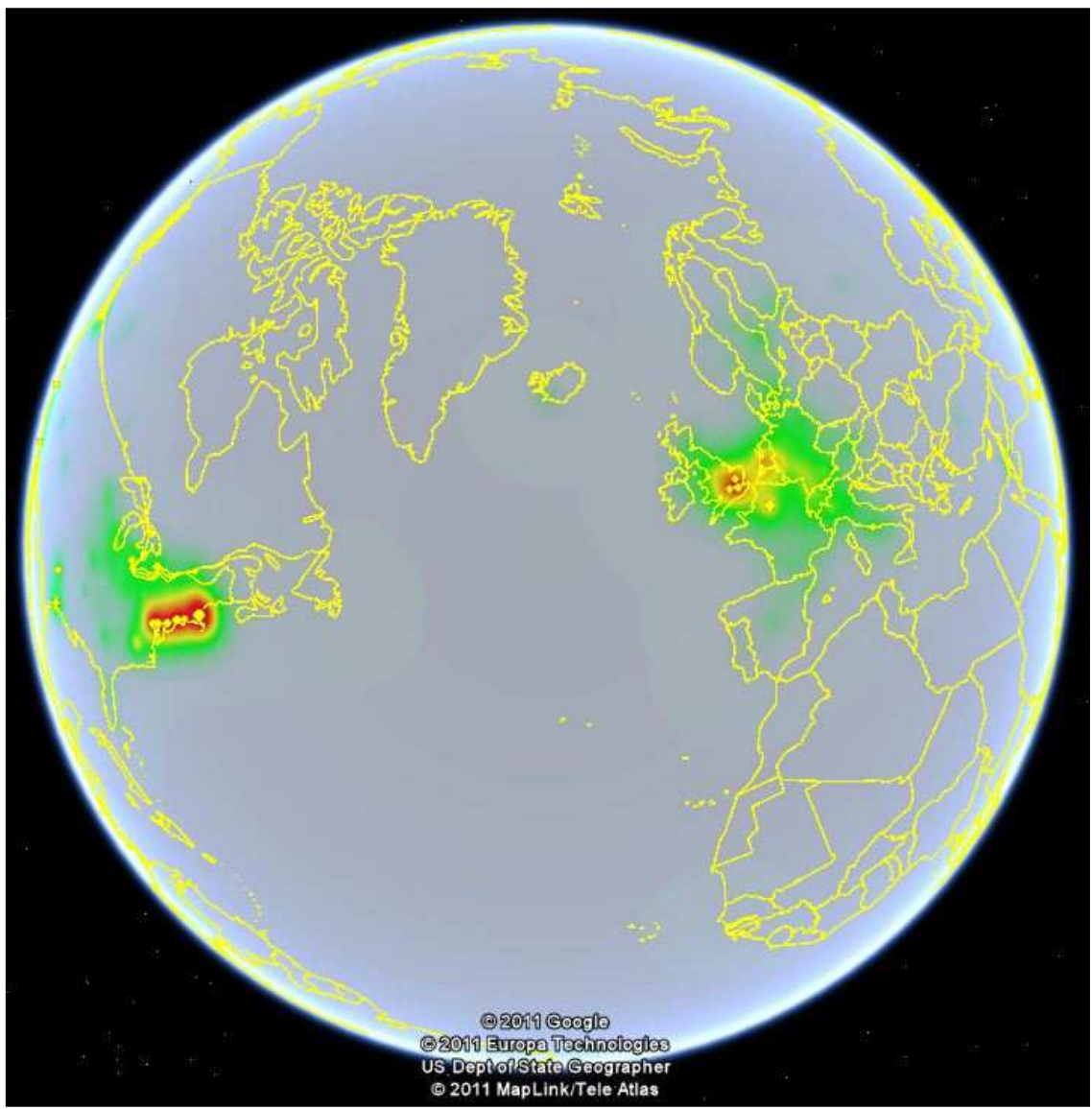

Figure 5. Density map of authors having published highly cited biochemistry, genetics & molecular biology papers in 2007 (searched in Scopus) by using Google Earth instead of traditional GIS software. This figure appears in color on the Web (PDF and HTML of this paper), but is not reproduced in color in the printed version.